\documentclass[pdflatex,sn-mathphys-num]{sn-jnl}

\usepackage{graphicx}
\usepackage{amsmath,amssymb,amsfonts}
\usepackage{amsthm}
\usepackage{mathtools}
\usepackage{booktabs}
\usepackage{xcolor}
\usepackage{seqsplit}
\hypersetup{pdftitle={Sharp bounds on the Eliashberg critical temperature},pdfauthor={Davis Rens},pdfkeywords={Eliashberg theory, superconductivity, critical temperature, comparison principle, spectral bounds, formal verification},bookmarksdepth=2}
\renewcommand{\pacsbullet}{\hbox{,\hskip2.5pt}}
\newcommand{\QzeroLow}{1.318140211994}
\newcommand{\RhoZero}{0.000015596}
\newcommand{\GtwoUp}{0.010963}
\newcommand{\LamTwoFloat}{5.3485\times10^{-4}}

\newcommand{\RegOneRho}{2.96523}

\newcommand{\RegOneR}{1.12441}
\newcommand{\RegOneCrit}{0.71189}
\newcommand{\RegTwoCells}{48}
\newcommand{\RegTwoN}{160}
\newcommand{\RegTwoRatio}{1.06}
\newcommand{\RegTwoWorstMargin}{3.96}
\newcommand{\RegTwoWorstLo}{1/4}
\newcommand{\RegTwoWorstHi}{277873/1048576}

\newcommand{\RegThreeVal}{1.084348}
\newcommand{\RegThreeVb}{4}
\newcommand{\ArbPrec}{200}
\newcommand{\CertSha}{cd41aff1b564efe08fa23eea1849b709fe1324661d20b26d55652bcd15df4d9d}
\newcommand{\ReceiptSha}{9ce805079f1f3a1abce66b757210adba40ee647f656fddf4588387664d893bfe}
\newcommand{\PlanSha}{197786ca6b2fb0676ffde142bf6c7bbec1b5c97f8e43d988bfcf53bd2decbaec}
\newcommand{\FlintVer}{0.9.0}
\newcommand{\CertRuntime}{2.1}
\newcommand{\IndepSha}{4a502d96d2cc311ba5948e67ebd861b36e913650778af7ab1fc46192c86c3e03}
\newcommand{\IndepScriptSha}{7242083ff94a7c1f939a39ea6df55f18b4398ae220fd7b87b9c28646356f8036}
\newcommand{\IndepN}{40}
\newcommand{\IndepVa}{1/8}
\newcommand{\IndepVb}{5}
\newcommand{\IndepRatio}{5/4}
\newcommand{\IndepCells}{17}
\newcommand{\IndepQ}{659/500}
\newcommand{\IndepKappa}{0.8048}
\newcommand{\IndepMidGap}{0.00826}
\newcommand{\IndepLargeGap}{0.4353}
\newcommand{\IndepRuntime}{0.25}

\theoremstyle{thmstyleone}
\newtheorem{theorem}{Theorem}
\newtheorem{lemma}[theorem]{Lemma}

\newtheorem{corollary}[theorem]{Corollary}
\theoremstyle{thmstyletwo}
\newtheorem{remark}[theorem]{Remark}
\theoremstyle{thmstylethree}
\newtheorem{definition}[theorem]{Definition}

\newcommand{\Sp}{\operatorname{sup\,spec}}
\newcommand{\vp}{\varpi}
\newcommand{\ip}[2]{\langle #1,#2\rangle}
\newcommand{\NN}{\mathbb{N}}
\newcommand{\RR}{\mathbb{R}}
\newcommand{\Pt}{\mathcal{P}_2}
\newcommand{\cone}{\mathcal{C}}
\newcommand{\hash}[1]{{\small\texttt{\expandafter\seqsplit\expandafter{#1}}}}
\newcommand{\lean}[1]{\texttt{#1}}

\begin{document}

\title[Sharp bounds on the Eliashberg critical temperature]{Sharp bounds on the Eliashberg critical temperature: a comparison principle, convexity in the spectrum, and the Kiessling--Altshuler--Yuzbashyan conjectures}

\author*{\fnm{Davis} \sur{Rens}}\email{davis@supermatics.io}

\affil*{\orgname{SuperMatics Inc.}, \orgaddress{\city{Berkeley}, \state{CA}, \country{USA}}}

\abstract{\unboldmath In Eliashberg theory of phonon-mediated superconductivity with vanishing Coulomb pseudopotential, the critical temperature $T_c$ marks the linear instability of the normal state. Kiessling, Altshuler and Yuzbashyan (KAY) characterized it through the top eigenvalue of a compact operator on Matsubara-frequency sequences. They proved that $T_c$ is unique at strong coupling and conjectured $T_c<C(\lambda\langle\omega^2\rangle)^{1/2}$ for every coupling $\lambda$, with $\langle\omega^2\rangle$ the second moment of the normalized phonon spectrum and $C\approx0.1827$ sharp, a bound Leavens stated in 1977 on partly numerical grounds. A preprint by Kondratev, not yet peer reviewed, gives a proof of this bound and of all-coupling uniqueness. We give an independent proof, for every spectrum with finite second moment, based on a comparison principle that is also Kondratev's central step and was proved to first order by Bergmann and Rainer: strengthening the pairing interaction at every Matsubara transfer, strictly at the smallest one, strictly raises the top eigenvalue. At fixed $\lambda$ and $\langle\omega^2\rangle$ a single Einstein mode maximizes $T_c$. Convexity of the top eigenvalue in the spectrum gives sharp bounds on $T_c$ by the moments $\int\omega^n\alpha^2F\,d\omega$, $0\le n\le1$, including Leavens's zeroth-moment bound $T_c\le c_0\int\alpha^2F\,d\omega$, $c_0\approx0.231$ numerically, optimal at moderate coupling. All bounds persist under a Coulomb pseudopotential with symmetric cutoff. A prefix decomposition of non-increasing vectors proves a third KAY conjecture, not treated in the preprint. The finite-dimensional core, including that conjecture, is formalized in Lean~4, and two computer-assisted proofs independent of the comparison principle confirm the Einstein bound.}



\maketitle

\section{Introduction}\label{sec:intro}

\subsection{Background}

Eliashberg theory~\cite{Eliashberg1960} is the standard framework for computing the critical temperature $T_c$ of phonon-mediated superconductors~\cite{Carbotte1990,Marsiglio2020}. Its input is the Eliashberg spectral function $\alpha^2F(\omega)$, which measures how strongly electrons couple to lattice vibrations of frequency $\omega$. Two numbers summarize it: the dimensionless coupling constant $\lambda=2\int_0^\infty\alpha^2F(\omega)\,\omega^{-1}d\omega$, and the second moment $\langle\omega^2\rangle$ of the normalized spectral measure $P(d\omega)=2\alpha^2F(\omega)/(\lambda\omega)\,d\omega$.

For weak coupling $T_c$ is exponentially small in $1/\lambda$. For strong coupling, Allen and Dynes~\cite{AllenDynes1975} found that it grows without bound but only like a square root,
\begin{equation}\label{eq:AD}
T_c\simeq C\sqrt{\lambda\langle\omega^2\rangle}\qquad(\lambda\to\infty),\qquad C\approx0.18,
\end{equation}
in units with $\hbar=k_B=1$. The right-hand side of~\eqref{eq:AD} has long been regarded as an upper bound on $T_c$ at every coupling. Moussa and Cohen state it as one, $T_c\le0.183\sqrt{\lambda\langle\omega^2\rangle}$ for $\mu^*=0$, citing Allen and Dynes and Leavens~\cite{MoussaCohen2006,AllenDynes1975,Leavens1977}; see also~\cite{SemenokAltshulerYuzbashyan2025}. Allen and Dynes also argued that a single sharp phonon line (an Einstein spectrum) maximizes $T_c$ at fixed $\lambda$ and $\langle\omega^2\rangle$: they proved that the Einstein spectrum maximizes each coefficient $\lambda(n)$ of their linearized gap equation at fixed $\lambda$ and $\langle\omega^2\rangle$, and noted that the top eigenvalue, and with it $T_c$, is maximized by making all $\lambda(n)$ as large as possible~\cite[Sect.~II and App.~A]{AllenDynes1975}. Bergmann and Rainer had shown that for $\mu^*=0$ added spectral weight at any frequency raises $T_c$ to first order, that is, that the functional derivative $\delta T_c/\delta\alpha^2F(\omega)$ is positive~\cite{BergmannRainer1973}.

Leavens stated the bound itself as a rigorous least upper bound for $\mu^*=0$, in the form $T_c\le0.258\,\big(\int_0^\infty\omega\,\alpha^2F(\omega)\,d\omega\big)^{1/2}$~\cite[eq.~(20)]{Leavens1977}. The first moment equals $\lambda\langle\omega^2\rangle/2$, so his constant is $0.258/\sqrt2\approx0.182$ in the units of~\eqref{eq:AD}. His derivation has two steps. A first-order variational argument based on the positivity of Bergmann and Rainer shows that a spectrum maximizing $T_c$ at fixed first moment can only be a single Einstein mode. This presupposes that a maximizer exists, whereas at fixed first moment the supremum is approached only by a mode of diverging weight at vanishing frequency, as he notes. The maximum over Einstein spectra is then read off from numerical solutions of the Eliashberg equations together with the strong-coupling asymptote~\cite[Sect.~2 and Fig.~1]{Leavens1977}; this second step is the content of Conjecture~1 of Part~III below. Two years earlier he had stated, as rigorously shown, the zeroth-moment bound $T_c\le0.2309\int_0^\infty\alpha^2F(\omega)\,d\omega$, with equality for the Einstein spectrum at frequency $1.750\int_0^\infty\alpha^2F(\omega)\,d\omega$, that is at coupling $\lambda=1.143$~\cite[eqs.~(13)--(14)]{Leavens1975}. Its derivation has the same two steps: the variational argument, which presupposes an optimal spectrum, followed by a numerical search over Einstein spectra~\cite[Sect.~2]{Leavens1975}; the 1977 paper restates the bound~\cite[eqs.~(18)--(19)]{Leavens1977}. In both papers Leavens argued that the bounds persist for $\mu^*>0$ because the Coulomb repulsion lowers $T_c$~\cite[Sect.~3]{Leavens1975,Leavens1977}.

In a series of three papers~\cite{KAY2025I,KAY2025II,KAY2025III}, Kiessling, Altshuler and Yuzbashyan (KAY) made these questions precise. Building on the spin-chain formulation of Yuzbashyan and Altshuler~\cite{YuzbashyanAltshuler2022SpinChain}, they showed that the normal state of the linearized, isotropic Eliashberg model with Coulomb pseudopotential $\mu^*=0$ is linearly unstable precisely when $\lambda$ times the top eigenvalue of an explicit compact self-adjoint operator exceeds one. They identified the constant in~\eqref{eq:AD} as $C=\sqrt{g(2)}/(2\pi)=0.1827262477\ldots$, where $g(2)$ is the top eigenvalue of the operator of their $\gamma$-model at $\gamma=2$~\cite{KAY2025I}, and they stated the following conjectures.
\begin{itemize}
\item \emph{Conjecture~1 of Part~III}~\cite{KAY2025III} (Einstein phonons of frequency $\Omega$): for all $\Omega>0$ and $\lambda>0$ there is a critical temperature $T_c(\lambda,\Omega)>0$, and $T_c(\lambda,\Omega)<\frac{\Omega}{2\pi}\sqrt{g(2)\lambda}$.
\item \emph{Conjecture~2 of Part~II}~\cite{KAY2025II} (general spectra $P$): for all admissible $P$ and $\lambda>0$ there is a critical temperature $T_c(\lambda,P)>0$, and $T_c(\lambda,P)<\frac{1}{2\pi}\sqrt{g(2)\lambda\langle\omega^2\rangle}$.
\item \emph{Conjecture~3 of Part~II}~\cite{KAY2025II}: a sharp lower bound, attained exactly by constant sequences, for a finite-dimensional Rayleigh quotient on non-negative non-increasing sequences.
\end{itemize}
KAY proved that $T_c$ is unique when $\lambda$ exceeds a spectrum-dependent threshold (for Einstein phonons, $\lambda>0.7670\ldots$~\cite[Thm.~5]{KAY2025III}), that $T_c/\sqrt{\lambda\langle\omega^2\rangle}$ has the limit $\sqrt{g(2)}/2\pi$ as $\lambda\to\infty$, and an explicit upper bound on $T_c$ whose constant is about $2.03$ times the conjectured one~\cite{KAY2025II,KAY2025III}. They noted that a proof of uniqueness for all $\lambda>0$, which the thermodynamic picture of the transition presupposes, would be mathematically desirable~\cite{KAY2025II,KAY2025III}. Their monotonicity argument needs the dimensionless phonon frequency $\vp=\Omega/(2\pi T)$ to satisfy $\vp\le\sqrt2$~\cite[Lemma~3]{KAY2025II}, and they remarked that proving monotonicity for all $\vp$ ``would require a different strategy''~\cite{KAY2025III}. Elezaby and Abanov sharpened rigorous bounds for the $\gamma$-model~\cite{ElezabyAbanov2026}.

In September 2026 Kondratev posted a preprint~\cite{Kondratev2026}, not yet peer reviewed, that gives a proof of Conjecture~2 of Part~II, and with it Conjecture~1 of Part~III, for every non-zero phonon measure with finite coupling and second moment. It shows that the top eigenvalue decreases strictly with temperature on all of $(0,\infty)$, so that the onset temperature is unique for every $\lambda>0$, and it also derives the optimality of Einstein spectra at fixed $\lambda$ and $\langle\omega^2\rangle$~\cite[Sect.~12]{Kondratev2026}. Its central step, a coefficient-by-coefficient comparison of two kernels on the ordered top eigenvector~\cite[Prop.~1]{Kondratev2026}, is in substance our Theorem~\ref{thm:comp}. We found the argument presented here without knowledge of that preprint and learned of it while preparing this manuscript. Section~\ref{sec:relation} compares the two treatments.

\subsection{Results}

We prove all three conjectures, together with uniqueness of $T_c$ at every coupling (Theorems~\ref{thm:bound},~\ref{thm:conj3} and~\ref{thm:unique}). Results~1--3 below are also obtained in~\cite{Kondratev2026}; results~4 and~5 are not treated there. Result~2 is the bound stated by Leavens, whose derivation is described above~\cite{Leavens1977}, and result~3 was argued by Allen and Dynes, with the monotonicity of the top eigenvalue in the coefficients read off from the equations rather than proved~\cite{AllenDynes1975}.
\begin{enumerate}
\item \emph{Uniqueness.} For every phonon spectrum with finite second moment and every $\lambda>0$ there is exactly one critical temperature: $\lambda k(P,T)>1$ for $T<T_c$ and $\lambda k(P,T)<1$ for $T>T_c$, so for KAY-admissible $P$ the linearized problem has a superconducting instability below $T_c$ and none above. $T_c$ increases continuously and strictly with $\lambda$.
\item \emph{The bound.} $T_c<\frac{1}{2\pi}\sqrt{g(2)\lambda\langle\omega^2\rangle}\approx0.1827\sqrt{\lambda\langle\omega^2\rangle}$ for every $\lambda>0$ and every spectrum. The constant cannot be improved.
\item \emph{Einstein spectra are optimal.} At fixed $\lambda$ and $\langle\omega^2\rangle$, a single Einstein mode at frequency $\langle\omega^2\rangle^{1/2}$ has a strictly higher $T_c$ than any other spectrum.
\item \emph{Conjecture~3 of Part~II} holds.
\item \emph{Moment bounds.} For $0\le n\le1$, $T_c\le\frac1{2\pi}\big(2H_n\int_0^\infty\omega^n\alpha^2F(\omega)\,d\omega\big)^{1/(n+1)}$, where $H_n$ is a constant defined by the Einstein problem, and the bound is the supremum of $T_c$ over Einstein spectra with the same moment (Theorem~\ref{thm:moments}). For $n=1$ this is result~2; for $n=0$ it is Leavens's bound $T_c\le(H_0/\pi)\int_0^\infty\alpha^2F(\omega)\,d\omega$~\cite{Leavens1975}, where $H_0/\pi\approx0.2309$ in floating point.
\end{enumerate}
The first three follow from a single statement, the comparison principle of Theorem~\ref{thm:comp}: if the pairing interaction is increased at every Matsubara frequency transfer, and strictly at the smallest transfer, then the top eigenvalue strictly increases. Conjecture~3 is proved separately in Section~\ref{sec:proofconj3}, by decomposing non-increasing vectors into prefix indicators. Result~5 rests on a second, elementary observation: the top eigenvalue is a supremum of functionals that are linear in the spectral measure, hence convex in it, so it is at most the average of the Einstein eigenvalues over the spectrum (Lemma~\ref{lem:jensen}), and at a fixed moment no spectrum does better than the supremum over Einstein spectra. This makes Leavens's reduction to Einstein spectra rigorous without assuming that a maximizing spectrum exists. Combined with the computer-assisted proofs of the Einstein inequality, it also proves that the normal state is stable above the bound of result~2 without using the comparison principle. Uniqueness also yields these bounds for every instability temperature when a Coulomb pseudopotential with a symmetric frequency cutoff is added (Corollary~\ref{cor:coulomb}), which makes Leavens's remark on $\mu^*>0$ rigorous in that model. Apart from this corollary, these are theorems about the linearized, isotropic Eliashberg model with $\mu^*=0$; Section~\ref{sec:scope} states what they do not cover.

\subsection{The idea of the proof}

The interaction between Matsubara frequencies $\omega_n$ and $\omega_m$ depends only on the transfer $|n-m|$, so the KAY operator is a sum $K=\sum_{j\ge1}a_jM_j$: one fixed operator $M_j$ per transfer $j$, weighted by the interaction strength $a_j$ at that transfer. Heuristically, raising $a_j$ changes the top eigenvalue, to first order, by $\ip{\xi}{M_j\xi}$, where $\xi$ is the optimizer. Each $M_j$ has two competing parts (Fig.~\ref{fig:pieces}): a pairing part, which couples frequencies a distance $j$ apart, directly or after reflecting one of them to $-\omega_m$, and a negative diagonal part, which comes from the mass renormalization $Z_n$ and acts as pair breaking. On a general vector either part can win. On the optimizer the pairing part always wins: Lemma~\ref{lem:mu} writes $\ip{\xi}{M_j\xi}$ as a sum of terms $y_{n+j}(y_n-y_{n+j})$ and $y_ny_m$, where $y_n\propto\Delta_n/\omega_n$ is the gap divided by the frequency, and Lemma~\ref{lem:cone} shows that $y$ is non-negative and non-increasing. Every term is then non-negative.

KAY instead differentiated with respect to $\vp^2$, grouping the transfers with weights that decrease in $j$ only when $\vp\le\sqrt2$~\cite[Lemma~3]{KAY2025II}. The transfer decomposition removes this restriction. For the $\gamma$-model, KAY had already expanded their Rayleigh quotient over transfers, with coefficients that are non-negative on non-increasing vectors, to prove that the quotient is positive~\cite[eqs.~(72)--(73)]{KAY2025II}; here the same expansion is used to compare different kernels. The first-order version of the whole argument goes back to Bergmann and Rainer~\cite[App.~A]{BergmannRainer1973}. Working with the pair-breaking eigenvalue of the linearized equations at a finite frequency cutoff, they obtained $\delta T_c/\delta\alpha^2F(\omega)$ from the Hellmann--Feynman theorem and showed that the coefficient of every transfer in it is non-negative, using that for $\mu^*=0$ the gap function is positive and decreasing in $|\omega_n|$. Theorem~\ref{thm:comp} is the corresponding statement for finite changes of the kernel: it needs no cutoff, no differentiability of the top eigenvalue and no monotonicity of the larger kernel, and the ordering of the optimizer is proved in full (Lemma~\ref{lem:cone}). It also supplies the step in the Allen--Dynes argument for Einstein optimality that they read off from their equations: the top eigenvalue increases when every $\lambda(n)$ does, although the pair-breaking diagonal terms grow with them. Likewise, Theorem~\ref{thm:bound} replaces both steps of Leavens's derivation: it needs no maximizing spectrum, and the Einstein inequality that he read off from numerical solutions is proved.

\subsection{Relation to Kondratev's preprint}\label{sec:relation}

The proof of the bound in~\cite{Kondratev2026} and the one given here have the same structure. The expansion of the quadratic form over transfers~\cite[eqs.~(9)--(10)]{Kondratev2026} is our Lemma~\ref{lem:mu}, the kernel comparison on the ordered top eigenvector~\cite[Prop.~1]{Kondratev2026} is our Theorem~\ref{thm:comp}, and both treatments obtain Einstein optimality from Jensen's inequality, as Allen and Dynes did for the coefficients $\lambda(n)$~\cite[App.~A]{AllenDynes1975}. Both rest on the transfer-by-transfer positivity that Bergmann and Rainer used at first order~\cite{BergmannRainer1973}. The two treatments differ in four respects. First, the ordering of the top eigenvector is proved there by Perron--Frobenius theory and power iteration, and here by an extremal argument that avoids the Perron--Frobenius theorem (Lemma~\ref{lem:cone}). Second, the passage to infinitely many frequencies uses norm convergence of the truncated eigenvectors there and weak compactness here. Third, the preprint encloses the constant, $1.318139<g<1.31815$, in exact arithmetic, whereas our proofs use $g$ only through its definition as a top eigenvalue; our certificates instead prove the Einstein inequality $h(\vp)<g\vp^2$ itself without the comparison principle (Appendix~\ref{app:cap}). Fourth, the preprint proves sharpness along a family of Einstein spectra, while Theorem~\ref{thm:bound}(c) also gives the limit $\lambda\to\infty$ for each fixed spectrum with finite fourth moment. Conjecture~3 of Part~II, the moment bounds of Theorem~\ref{thm:moments}, Corollary~\ref{cor:coulomb} and the formalization of Section~\ref{sec:lean} have no counterpart in~\cite{Kondratev2026}.

\subsection{Verification and organization}

For every finite truncation, the cone lemma, transfer decomposition, comparison principle, Einstein bound, monotonicity and non-strict Einstein optimality for discrete spectra are formalized and machine-checked in Lean~4 with Mathlib~\cite{Lean4,Mathlib2020}, as is Conjecture~3 in full (Section~\ref{sec:lean}). The text supplies the compactness argument needed for infinitely many frequencies. Two further computer-assisted proofs of the Einstein bound, independent of the comparison principle, use interval and exact rational arithmetic (Appendix~\ref{app:cap}).

Section~\ref{sec:setting} derives the KAY operator from the linearized Eliashberg equations and collects its basic properties. Section~\ref{sec:results} states the results. Sections~\ref{sec:proofcomp}--\ref{sec:proofconj3} contain the proofs. Section~\ref{sec:lean} describes the formalization and Section~\ref{sec:scope} the scope of the results and some open problems.

\section{The linearized Eliashberg problem}\label{sec:setting}

\subsection{From the Eliashberg equations to an eigenvalue problem}

Let $T>0$ and $\omega_n=\pi T(2n+1)$, $n\in\mathbb Z$, be the fermionic Matsubara frequencies. The isotropic Eliashberg equations with $\mu^*=0$, linearized in the gap function $\Delta_n$, read~\cite{AllenMitrovic1982,Carbotte1990,Marsiglio2020}
\begin{align}
Z_n\,\omega_n&=\omega_n+\pi T\sum_{m\in\mathbb Z}\lambda(\omega_n-\omega_m)\,\operatorname{sgn}\omega_m,\label{eq:Z}\\
Z_n\,\Delta_n&=\pi T\sum_{m\in\mathbb Z}\lambda(\omega_n-\omega_m)\,\frac{\Delta_m}{|\omega_m|},\label{eq:gap}
\end{align}
with the interaction
\[
\lambda(\nu)=2\int_0^\infty\frac{\omega\,\alpha^2F(\omega)}{\omega^2+\nu^2}\,d\omega=\lambda\int_0^\infty\frac{\omega^2}{\omega^2+\nu^2}\,P(d\omega),\qquad\lambda=\lambda(0).
\]
Since $\omega_n-\omega_m=2\pi T(n-m)$, the interaction depends only on the transfer $|n-m|$. We write
\begin{equation}\label{eq:kernel}
\lambda(\omega_n-\omega_m)=\lambda\,a_{|n-m|},\qquad a_j=a_j(P,T):=\int_0^\infty\frac{\omega^2}{\omega^2+(2\pi Tj)^2}\,P(d\omega),
\end{equation}
so that $a_0=1$ and $1>a_1\ge a_2\ge\dots>0$. Put $w_n:=2n+1$ and $A_n:=\sum_{j=1}^na_j$.

We look for even solutions, $\Delta_{-n-1}=\Delta_n$, and restrict to $n\ge0$. Splitting the sum in~\eqref{eq:Z} into $m\ge0$ and $m<0$ and telescoping gives $\sum_{m\in\mathbb Z}a_{|n-m|}\operatorname{sgn}\omega_m=\sum_{m\ge0}(a_{|n-m|}-a_{n+m+1})=1+2A_n$, hence
\begin{equation}\label{eq:Zsol}
Z_n=1+\frac{\lambda\,(1+2A_n)}{w_n}.
\end{equation}
In~\eqref{eq:gap} the same splitting gives $Z_n\Delta_n=\lambda\sum_{m\ge0}\big(a_{|n-m|}+a_{n+m+1}\big)\Delta_m/w_m$. Insert~\eqref{eq:Zsol} and set $y_n:=\Delta_n/w_n$. The term $m=n$ of the first sum contributes $\lambda a_0y_n=\lambda y_n$ on the right, which cancels the term $\lambda y_n$ coming from the $1$ in $1+2A_n$ on the left. What remains is
\[
w_ny_n=\lambda\Big(\sum_{m\ge0}\big[(1-\delta_{nm})a_{|n-m|}+a_{n+m+1}\big]y_m-2A_ny_n\Big).
\]
Dividing by $\sqrt{w_n}$ and writing $\xi_n:=\sqrt{w_n}\,y_n=\Delta_n/\sqrt{w_n}$ turns this into the symmetric eigenvalue problem
\begin{equation}\label{eq:eig}
\xi=\lambda\,K[a]\,\xi,
\end{equation}
where, for any summable sequence $a=(a_j)_{j\ge1}$, $K[a]$ is the operator on real $\ell^2(\NN_0)$ with matrix
\begin{equation}\label{eq:K}
K[a]_{nm}=\frac{(1-\delta_{nm})\,a_{|n-m|}+a_{n+m+1}}{\sqrt{w_nw_m}}-\delta_{nm}\,\frac{2A_n}{w_n}.
\end{equation}
The value $a_0$ does not enter. This is the operator $K(P,T)$ of KAY~\cite[eqs.~(39)--(42)]{KAY2025II}, whose quadratic form $\ip{\xi}{(I-\lambda K)\xi}$ agrees, up to a positive overall factor, with the second variation of their spin-chain free energy at the normal state. KAY proved~\cite[Thms.~1, 1$^+$ and~3]{KAY2025II} that the normal state is linearly unstable against superconducting perturbations if and only if $\lambda\,k(P,T)>1$, where
\[
k(P,T):=\Sp K[a(P,T)]
\]
is the largest eigenvalue. The critical temperature is therefore the temperature at which $\lambda\,k(P,T)=1$.

In KAY's spin-chain formulation, $y_n=\xi_n/\sqrt{w_n}=\Delta_n/w_n$ is, to linear order, $\pi T$ times the spin angle $\theta_n$ of~\cite{KAY2025II,YuzbashyanAltshuler2022SpinChain}. It is proportional to $\Delta_n/\omega_n$ rather than to the gap itself. The condition $\xi\in\ell^2$ means $\sum_n\Delta_n^2/w_n<\infty$.

\subsection{Kernels}

\begin{definition}\label{def:kernel}
A \emph{kernel} is a sequence $a=(a_j)_{j\ge1}$ of positive numbers that is non-increasing in $j$ and summable.
\end{definition}

We use three families of kernels.
\begin{itemize}
\item \emph{General spectra.} Let $\Pt$ be the set of probability measures on $(0,\infty)$ with $\langle\omega^2\rangle:=\int\omega^2P(d\omega)<\infty$. For $P\in\Pt$ and $T>0$, $a(P,T)$ in~\eqref{eq:kernel} is a kernel, since $a_j(P,T)<\langle\omega^2\rangle/(2\pi Tj)^2$. KAY's admissible class~\cite{KAY2025II}, spectra supported in $[0,\Omega(P)]$ with density proportional to $\omega$ near $0$, is contained in $\Pt$.
\item \emph{Einstein phonons.} For $P=\delta_\Omega$ the kernel depends only on $\vp:=\Omega/(2\pi T)$, namely $a_j=\vp^2/(\vp^2+j^2)$. We write $h(\vp):=\Sp K[(\vp^2/(\vp^2+j^2))_j]$; this is the top eigenvalue of KAY's Einstein operator~\cite{KAY2025III}.
\item \emph{The $\gamma$-model at $\gamma=2$.} The kernel $(j^{-2})_{j\ge1}$ gives KAY's $\gamma$-model operator $G_0:=K[(j^{-2})_j]$ at $\gamma=2$~\cite{KAY2025I}, and
\[
g:=g(2)=\Sp G_0=1.31814\ldots
\]
This operator describes the strong-coupling regime: as $T\to\infty$ one has $a_j(P,T)\approx\langle\omega^2\rangle(2\pi T)^{-2}j^{-2}$, hence $k(P,T)\approx g\langle\omega^2\rangle/(2\pi T)^2$.
\end{itemize}

\subsection{Basic properties of the operator}

\begin{figure}[t]
\centering
\includegraphics[width=\textwidth]{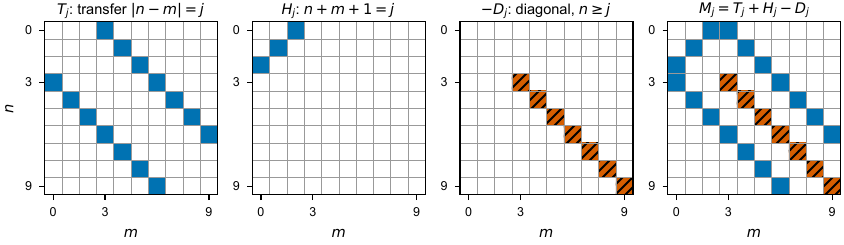}
\caption{Sign pattern of the transfer operator $M_j$ in~\eqref{eq:Mj} for $j=3$ on the first ten Matsubara frequencies. The panels show the pairing terms $T_j$ and $H_j$, the diagonal term $-D_j$, and their sum. Positive entries are solid blue and negative entries are hatched orange. The operator $M_j$ is indefinite, but Lemma~\ref{lem:mu} shows that it is non-negative on the cone containing the optimizer.}
\label{fig:pieces}
\end{figure}

Write $[\cdot]$ for the indicator of a statement. Every $K[a]$ is a weighted sum of fixed operators, one per transfer:
\begin{gather}
K[a]=\sum_{j\ge1}a_jM_j,\qquad M_j=T_j+H_j-D_j,\label{eq:Mj}\\
(T_j)_{nm}=\frac{[\,|n-m|=j\,]}{\sqrt{w_nw_m}},\quad(H_j)_{nm}=\frac{[\,n+m+1=j\,]}{\sqrt{w_nw_m}},\quad(D_j)_{nm}=\delta_{nm}\,\frac{2\,[n\ge j]}{w_n}.\nonumber
\end{gather}
Here $T_j$ (Toeplitz) couples frequencies at distance $j$, $H_j$ (Hankel) couples $\omega_n$ to the reflected frequency $-\omega_m$ at distance $j$, and $D_j$ is the contribution of the transfer $j$ to the mass renormalization~\eqref{eq:Zsol}: $a_j$ enters $Z_n$ for every $n\ge j$. Fig.~\ref{fig:pieces} shows the pattern.

\begin{lemma}\label{lem:basic}
Let $a,b$ be summable sequences of real numbers.
\begin{enumerate}
\item[(a)] $\|T_j\|\le2/\sqrt{2j+1}$, $\|H_j\|\le1/\sqrt{2j-1}$ and $\|D_j\|=2/(2j+1)$; hence $\|M_j\|<3$, the series~\eqref{eq:Mj} converges in operator norm, and $\|K[a]-K[b]\|\le3\sum_j|a_j-b_j|$.
\item[(b)] $K[a]$ is compact and self-adjoint, and $K[sa]=sK[a]$ for $s\in\RR$.
\item[(c)] If $a_1>0$, then $\kappa[a]:=\Sp K[a]\ge K[a]_{00}=a_1>0$, so $\kappa[a]$ is the largest eigenvalue of $K[a]$.
\item[(d)] Let $K^{(N)}[a]$ be the $N\times N$ upper-left block of $K[a]$ (the truncation to the first $N$ positive Matsubara frequencies) and $\kappa^{(N)}[a]$ its largest eigenvalue. Then $\kappa^{(N)}[a]\le\kappa^{(N+1)}[a]$, and $\kappa^{(N)}[a]\to\kappa[a]$ as $N\to\infty$.
\end{enumerate}
\end{lemma}

\begin{proof}
(a) $T_j=S_j+S_j^{\mathsf T}$, where $S_j$ maps $e_{n+j}\mapsto e_n/\sqrt{w_nw_{n+j}}$ and has norm $\sup_n(w_nw_{n+j})^{-1/2}=(2j+1)^{-1/2}$. $H_j$ is a finite anti-diagonal matrix with entries $(w_nw_{j-1-n})^{-1/2}\le(2j-1)^{-1/2}$. $D_j$ is diagonal with largest entry $2/w_j$. For $j=1$ the three bounds sum to $2/\sqrt3+1+2/3<3$, and all three decrease in $j$. (b) $S_j$ is a weighted shift with weights tending to $0$, $H_j$ has finite rank and $D_j$ is diagonal with entries tending to $0$; so each $M_j$ is compact, and so is the norm-convergent sum. (c) $K[a]_{00}=a_1$ because $A_0=0$. A compact self-adjoint operator whose spectrum contains points above $0$ has its supremum as an eigenvalue. (d) $K^{(N)}$ is a compression of $K^{(N+1)}$, so $\kappa^{(N)}$ is non-decreasing in $N$ and $\kappa^{(N)}\le\Sp K$. Every finitely supported unit vector lies in some truncation, and such vectors are dense in the unit sphere; since the quadratic form of the bounded operator $K$ is continuous, $\sup_N\kappa^{(N)}[a]=\Sp K[a]$.
\end{proof}

\section{Main results}\label{sec:results}

\begin{theorem}[Comparison principle]\label{thm:comp}
Let $a$ be a kernel and $b=(b_j)_{j\ge1}$ a sequence with $b_j\ge a_j$ for every $j\ge1$, $b_1>a_1$, and $\sum_j(b_j-a_j)<\infty$. Then
\[
\kappa[b]>\kappa[a],\qquad\text{and}\qquad\kappa^{(N)}[b]>\kappa^{(N)}[a]\ \text{ for every }N\ge1.
\]
\end{theorem}

In Theorem~\ref{thm:comp} only $a$ needs to be non-increasing; $b$ need not be monotone. Monotonicity of $a$ is used only to locate the optimizer of $K[a]$ (Lemma~\ref{lem:cone}).

We write $k^{(N)}(P,T):=\kappa^{(N)}[a(P,T)]$ and $h^{(N)}(\vp):=\kappa^{(N)}[(\vp^2/(\vp^2+j^2))_j]$ for the corresponding truncated eigenvalues.

\begin{theorem}[Uniqueness of $T_c$ at every coupling]\label{thm:unique}\leavevmode
\begin{enumerate}
\item[(a)] The Einstein eigenvalue $\vp\mapsto h(\vp)$ is continuous and strictly increasing on $(0,\infty)$, with $h(0^+)=0$ and $h(\vp)\to\infty$ as $\vp\to\infty$.
\item[(b)] For every $P\in\Pt$, $T\mapsto k(P,T)$ is continuous and strictly decreasing on $(0,\infty)$, with $k(P,T)\to\infty$ as $T\to0$ and $k(P,T)\to0$ as $T\to\infty$.
\item[(c)] Consequently, for every $P\in\Pt$ and every $\lambda>0$ there is exactly one $T_c(\lambda,P)>0$ with $\lambda k(P,T_c)=1$. One has $\lambda k(P,T)>1$ for $T<T_c$ and $\lambda k(P,T)<1$ for $T>T_c$, so for KAY-admissible $P$ the normal state is linearly unstable exactly when $T<T_c$. The map $\lambda\mapsto T_c(\lambda,P)$ is a continuous, strictly increasing bijection of $(0,\infty)$ onto itself.
\item[(d)] For Einstein phonons of frequency $\Omega$, $T_c(\lambda,\Omega)=\Omega/\big(2\pi h^{-1}(1/\lambda)\big)$.
\end{enumerate}
For every truncation $N\ge1$, the maps $\vp\mapsto h^{(N)}(\vp)$ and $T\mapsto k^{(N)}(P,T)$ are likewise continuous and, respectively, strictly increasing and strictly decreasing.
\end{theorem}

\begin{theorem}[The sharp bound; Conjecture~1 of Part~III and Conjecture~2 of Part~II]\label{thm:bound}\leavevmode
\begin{enumerate}
\item[(a)] $h(\vp)<g\,\vp^2$ for every $\vp>0$, and $k(P,T)<g\,\langle\omega^2\rangle/(2\pi T)^2$ for every $P\in\Pt$ and $T>0$.
\item[(b)] Consequently, for every $\lambda>0$,
\[
T_c(\lambda,\Omega)<\frac{\Omega}{2\pi}\sqrt{g\lambda},\qquad T_c(\lambda,P)<\frac{1}{2\pi}\sqrt{g\lambda\langle\omega^2\rangle}=0.182726\ldots\sqrt{\lambda\langle\omega^2\rangle}.
\]
\item[(c)] The constant is sharp: $h(\vp)/(g\vp^2)\to1$ as $\vp\to0$, hence $T_c(\lambda,\Omega)\big/\big(\frac{\Omega}{2\pi}\sqrt{g\lambda}\big)\to1$ as $\lambda\to\infty$. The same limit holds for every $P\in\Pt$ with $\int\omega^4P(d\omega)<\infty$.
\end{enumerate}
\end{theorem}

\begin{theorem}[Einstein spectra are optimal]\label{thm:einstein}
Let $P\in\Pt$ and $\Omega_2:=\langle\omega^2\rangle^{1/2}$. Then $k(P,T)\le h(\Omega_2/2\pi T)$ for every $T>0$, with strict inequality unless $P=\delta_{\Omega_2}$. Hence $T_c(\lambda,P)<T_c(\lambda,\Omega_2)$ for every $\lambda>0$ unless $P=\delta_{\Omega_2}$.
\end{theorem}

\begin{figure}[t]
\centering
\includegraphics[width=0.49\textwidth]{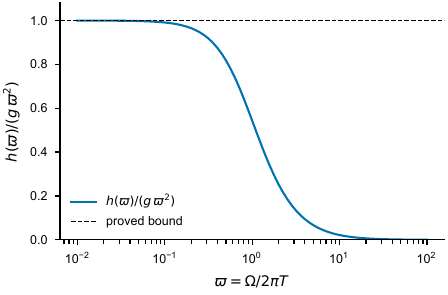}\hfill
\includegraphics[width=0.49\textwidth]{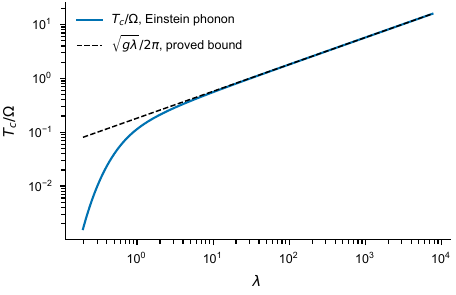}
\caption{Floating-point illustration for Einstein phonons, with truncation $N\ge800$. The curves are not used in the proofs. Left: $h(\vp)/(g\vp^2)$, which Theorem~\ref{thm:bound} proves is below $1$ and tends to $1$ as $\vp\to0$ (strong coupling). Right: $T_c/\Omega$ against $\lambda$, with the ceiling $\sqrt{g\lambda}/2\pi$.}
\label{fig:curves}
\end{figure}

Theorems~\ref{thm:unique} and~\ref{thm:bound} together prove Conjecture~1 of~\cite{KAY2025III} and Conjecture~2 of~\cite{KAY2025II}, which KAY stated for their admissible class; we obtain them, as does~\cite{Kondratev2026}, on the larger class $\Pt$. Theorem~\ref{thm:einstein} says that at fixed $\lambda$ and $\langle\omega^2\rangle$ the Einstein spectrum at frequency $\Omega_2=\langle\omega^2\rangle^{1/2}$ has the largest $T_c$. Fig.~\ref{fig:curves} illustrates the Einstein case.

\begin{remark}[Adding spectral weight]\label{rem:weight}
Theorem~\ref{thm:comp} also gives the global form of the result of Bergmann and Rainer~\cite{BergmannRainer1973}. Write $\nu(d\omega)=\lambda P(d\omega)=2\alpha^2F(\omega)\omega^{-1}d\omega$, so that $\lambda a_j(P,T)=\int\omega^2(\omega^2+(2\pi Tj)^2)^{-1}\nu(d\omega)$ and, since $K[a]$ is linear in $a$, $\lambda k(P,T)=\kappa[\lambda a(P,T)]$. Let $\nu'=\nu+\eta$, where $\eta\ne0$ is a finite measure on $(0,\infty)$ with $\int\omega^2\eta(d\omega)<\infty$, and let $\lambda'$ and $P'$ be the coupling and normalized spectrum of $\nu'$. At every $T>0$ each coefficient increases strictly, by at most $\int\omega^2\eta(d\omega)/(2\pi Tj)^2$, so Theorem~\ref{thm:comp} gives $\lambda'k(P',T)>\lambda k(P,T)$, and Theorem~\ref{thm:unique}(b) gives $T_c(\lambda',P')>T_c(\lambda,P)$. Adding phonon spectral weight at any frequencies strictly raises $T_c$.
\end{remark}

\begin{theorem}[Sharp moment bounds]\label{thm:moments}
For $0\le n\le1$ let
\[
M_n:=\int_0^\infty\omega^n\,\alpha^2F(\omega)\,d\omega=\frac{\lambda}{2}\int\omega^{n+1}P(d\omega),\qquad H_n:=\sup_{\vp>0}\frac{h(\vp)}{\vp^{\,n+1}} .
\]
Then $H_n\le\max\{g,3\pi/2\}$ and $H_1=g$. For every $P\in\Pt$, $\lambda>0$ and $T>0$,
\[
\lambda k(P,T)\le\frac{2H_nM_n}{(2\pi T)^{n+1}},\qquad\text{hence}\qquad T_c(\lambda,P)\le\frac{1}{2\pi}\big(2H_nM_n\big)^{1/(n+1)}.
\]
The constant is sharp: the right-hand side is the supremum of $T_c$ over Einstein spectra with the same $M_n$, the coupling $\lambda$ being free.
\end{theorem}

For $n=1$, $2M_1=\lambda\langle\omega^2\rangle$ and Theorem~\ref{thm:moments} is the bound of Theorem~\ref{thm:bound}, without strict inequality. For $n=0$ it is Leavens's zeroth-moment bound~\cite[eq.~(13)]{Leavens1975}. A floating-point evaluation of $h$ on truncations with $N=400$ and $N=800$, which agree to $10^{-10}$, gives $H_0\approx0.72529$, attained at $\vp_0\approx1.2056$, so that
\[
T_c\le\frac{H_0}{\pi}\int_0^\infty\alpha^2F(\omega)\,d\omega\approx0.23087\int_0^\infty\alpha^2F(\omega)\,d\omega ,
\]
with equality, to the accuracy of these values, for the Einstein spectrum of weight $M_0$ at frequency $\Omega=2\pi T_c\vp_0\approx1.749\,M_0$, that is $\lambda=2M_0/\Omega\approx1.144$, close to Leavens's numerical values $0.2309$, $1.750$ and $1.143$~\cite[Sect.~2]{Leavens1975}. These numerical values of $H_0$ and $\vp_0$ are not certified; the inequality holds with $H_0$ defined as the supremum. Unlike the first-moment bound, which is approached only as $\lambda\to\infty$, the zeroth-moment bound is, numerically, attained at moderate coupling, where Migdal--Eliashberg theory is usually applied. For $n>1$ no such bound exists, since $h(\vp)/(g\vp^2)\to1$ as $\vp\to0$ forces $H_n=\infty$, as Leavens observed~\cite[eq.~(17)]{Leavens1977}; for $-1<n<0$ finiteness of $H_n$ would need a growth bound on $h$ that we do not prove.

\begin{corollary}[Coulomb pseudopotential with a symmetric cutoff]\label{cor:coulomb}
Let $P\in\Pt$, $\lambda>0$, $\mu^*\ge0$ and $\omega_c>0$. Replace $\lambda(\omega_n-\omega_m)$ in~\eqref{eq:gap} by $\lambda(\omega_n-\omega_m)-\mu^*\,\mathbf 1\{|\omega_n|<\omega_c\}\,\mathbf 1\{|\omega_m|<\omega_c\}$ and leave~\eqref{eq:Z} unchanged. The cutoff is placed symmetrically on both frequencies, so that the Coulomb correction is self-adjoint; as in the usual pseudopotential treatment, the Coulomb term does not enter the mass renormalization~\cite{AllenDynes1975,Carbotte1990}. If at some $T>0$ the resulting equations have a non-zero even solution with $\xi\in\ell^2$, then
\[
T\le T_c(\lambda,P)<\frac{1}{2\pi}\sqrt{g\lambda\langle\omega^2\rangle},
\]
and $T$ obeys the bounds of Theorem~\ref{thm:moments}. The same holds for the truncation to the first $N$ positive frequencies, and for any repulsive term whose matrix in the variables $\xi$ is positive semidefinite.
\end{corollary}

\begin{proof}
For even solutions the Coulomb term adds $-2\mu^*\,\mathbf 1\{\omega_n<\omega_c\}\sum_{m\ge0,\ \omega_m<\omega_c}y_m$ to the right-hand side of the equation for $w_ny_n$ in Section~\ref{sec:setting}, so~\eqref{eq:eig} becomes $\xi=(\lambda K[a]-2\mu^*vv^{\mathsf T})\xi$ with $v_n=\mathbf 1\{\omega_n<\omega_c\}/\sqrt{w_n}$, a vector with finitely many non-zero entries. The inner product with $\xi$ gives $\lambda\ip{\xi}{K[a]\xi}=\|\xi\|^2+2\mu^*\ip{v}{\xi}^2\ge\|\xi\|^2$, hence $\lambda k(P,T)\ge1$. By Theorem~\ref{thm:unique}(c) this excludes $T>T_c(\lambda,P)$, and Theorem~\ref{thm:bound}(b) bounds $T_c(\lambda,P)$. For the truncation the same computation gives $\lambda k^{(N)}(P,T)\ge1$, and $k^{(N)}\le k$ by Lemma~\ref{lem:basic}(d). For a general positive semidefinite repulsion the inner product argument is unchanged.
\end{proof}

Corollary~\ref{cor:coulomb} makes rigorous, in this model, Leavens's remark that the bounds for $\mu^*=0$ also apply for $\mu^*>0$~\cite[Sect.~3]{Leavens1975,Leavens1977}. It says nothing about uniqueness of the instability temperature when $\mu^*>0$: the Coulomb term is negative and frequency independent, and the optimizer need not lie in the cone $\cone$.

The last result concerns the quadratic forms behind KAY's finite-dimensional analysis. For $N\ge1$, a sequence $a$ and $\theta\in\RR^N$ put
\begin{align*}
\mathcal G_N[a](\theta)&:=\sum_{n,m<N}\theta_n\big[(1-\delta_{nm})a_{|n-m|}+a_{n+m+1}\big]\theta_m-\sum_{n<N}2A_n\,\theta_n^2,\\
\mathcal D_N(\theta)&:=\sum_{n<N}w_n\,\theta_n^2 .
\end{align*}
With $\xi_n=\sqrt{w_n}\,\theta_n$ these are $\ip{\xi}{K^{(N)}[a]\xi}$ and $\|\xi\|^2$, so $\mathcal G_N/\mathcal D_N$ is the Rayleigh quotient of $K^{(N)}[a]$ written in the variables $\theta=y$, which are proportional to KAY's linearized spin angles (Section~\ref{sec:setting}). For $a_j=j^{-\gamma}$ they are the numerator and denominator of~\cite[l.h.s.\ of (70)]{KAY2025II}. KAY proved the quotient positive on non-zero, non-negative, non-increasing $\theta$, showed that constant sequences are local minimizers, and conjectured that they are global minimizers~\cite[Conjecture~3]{KAY2025II}.

\begin{theorem}[Conjecture~3 of Part~II]\label{thm:conj3}
Let $N\ge1$ and let $a$ be positive and strictly decreasing, for instance $a_j=j^{-\gamma}$ with $\gamma>0$. Put
\[
c_N[a]:=\frac{1}{N^2}\sum_{n,m<N}a_{n+m+1}=\frac{1}{N^2}\sum_{k=1}^{2N-1}\min\{k,2N-k\}\,a_k .
\]
Then for every non-zero, non-negative, non-increasing $\theta\in\RR^N$,
\[
\mathcal G_N[a](\theta)\ \ge\ c_N[a]\,\mathcal D_N(\theta),
\]
with equality if and only if $\theta$ is constant.
\end{theorem}

Since the optimizer of $K^{(N)}[a]$ lies in the same cone (Lemma~\ref{lem:cone}), Theorem~\ref{thm:conj3} says that on this cone the Rayleigh quotient ranges between $c_N[a]$, attained by constants, and $\kappa^{(N)}[a]$, attained by the optimizer.

\section{Proof of the comparison principle}\label{sec:proofcomp}

For $\xi\in\ell^2(\NN_0)$ write $y_n:=\xi_n/\sqrt{w_n}$, the gap divided by the frequency, up to normalization (Section~\ref{sec:setting}). Let
\[
\cone:=\{\xi\in\ell^2(\NN_0):\ \xi_n\ge0\text{ for all }n\text{, and }y\text{ is non-increasing}\}
\]
and let $\cone_N\subset\RR^N$ be its finite-dimensional analogue. In physical terms, $\xi\in\cone$ means $\Delta_n\ge0$ with $\Delta_n/\omega_n$ non-increasing.

\subsection{Each transfer operator is non-negative on the cone}

\begin{lemma}\label{lem:mu}
If $\xi\in\cone$, then for every $j\ge1$
\begin{equation}\label{eq:mu}
\mu_j(\xi):=\ip{\xi}{M_j\xi}=2\sum_{n\ge0}y_{n+j}\,(y_n-y_{n+j})+\sum_{n+m+1=j}y_ny_m\ \ge\ 0,
\end{equation}
and $\mu_1(\xi)\ge y_0^2$. The same holds for $\xi\in\cone_N$ and the truncated forms $\mu_j^{(N)}(\xi):=\ip{\xi}{K^{(N)}[e_j]\xi}$, with the sums restricted to indices below $N$.
\end{lemma}

\begin{proof}
From~\eqref{eq:Mj}, $\ip{\xi}{T_j\xi}=2\sum_ny_ny_{n+j}$, $\ip{\xi}{D_j\xi}=2\sum_{n\ge j}y_n^2=2\sum_ny_{n+j}^2$ and $\ip{\xi}{H_j\xi}=\sum_{n+m+1=j}y_ny_m$. All series converge absolutely because $\sum_ny_n^2\le\|\xi\|^2$. Subtracting gives~\eqref{eq:mu}. On $\cone$ every summand is non-negative, and for $j=1$ the Hankel sum is $y_0^2$. In the truncated case the Toeplitz and diagonal sums run over $n+j<N$, and the same pairing applies.
\end{proof}

In~\eqref{eq:mu}, the pair-breaking term $-2y_{n+j}^2$ is dominated by the pairing term $2y_ny_{n+j}$ because $y_n\ge y_{n+j}$. For $a_j=j^{-\gamma}$ and finite $N$, the sum $\sum_ja_j\mu^{(N)}_j$ is KAY's expansion~\cite[eqs.~(72)--(73)]{KAY2025II}; the same identity is~\cite[eqs.~(9)--(10)]{Kondratev2026}.

\subsection{The optimizer lies in the cone}

\begin{lemma}\label{lem:cone}
Let $a$ be a kernel. Then $K[a]$ has a unit eigenvector $\xi\in\cone$ for $\kappa[a]$, and $y_0>0$. The same holds for every truncation $K^{(N)}[a]$, with $\xi\in\cone_N$ an eigenvector for $\kappa^{(N)}[a]$ whose entries are all positive.
\end{lemma}

For finite $N$ this is~\cite[Prop.~8]{KAY2025II}. We give a self-contained proof that avoids the Perron--Frobenius theorem; it is the proof formalized in Lean (Section~\ref{sec:lean}).

\begin{proof}
\emph{Finite $N$.} Write $K=K^{(N)}[a]$ and $\kappa=\kappa^{(N)}[a]$; the case $N=1$ is trivial, so let $N\ge2$. The unit sphere is compact, so the quadratic form $\ip{x}{Kx}$ attains its maximum $\kappa$ on it. Since all off-diagonal entries of $K$ are positive, $\ip{|x|}{K|x|}\ge\ip{x}{Kx}$, so there is a maximizer $z\ge0$. A maximizer is an eigenvector: the function $t\mapsto\ip{z+tu}{K(z+tu)}-\kappa\|z+tu\|^2$ is non-positive and vanishes at $t=0$, so its derivative there, $2\ip{u}{Kz-\kappa z}$, vanishes for every $u$. If $z_n=0$ for some $n$, the $n$-th row of $Kz=\kappa z$ reads $\sum_{m\ne n}K_{nm}z_m=0$, which forces $z=0$; hence $z_n>0$ for all $n$.

Put $y_n:=z_n/\sqrt{w_n}>0$, $a_0:=a_1$ and $C_{nm}:=a_{|n-m|}+a_{n+m+1}$ for $n,m<N$. Multiplying the $n$-th row of $Kz=\kappa z$ by $\sqrt{w_n}$ and adding $(a_0+2A_n)y_n$ to both sides gives
\[
(Cy)_n=d_n\,y_n,\qquad d_n:=\kappa\,w_n+a_0+2A_n>0,
\]
and $d_n$ is increasing in $n$. So the matrix $Q:=\operatorname{diag}(d)^{-1}C$ has positive entries and satisfies $Qy=y$.

$Q$ maps the cone $\cone'_N$ of non-negative non-increasing vectors into itself. The cone consists of non-negative combinations of the prefix indicators $u_l$, $1\le l\le N$, where $u_l$ is the indicator of $\{0,\dots,l-1\}$, and
\[
(Cu_l)_n=\sum_{i=n-l+1}^{n+l}a_{|i|}=:F_l(n),\qquad F_l(n+1)-F_l(n)=a_{n+l+1}-a_{|n-l+1|}\le0,
\]
because $n+l+1>|n-l+1|$ and $a$ is non-increasing on $\NN_0$ (recall $a_0=a_1$). So $C\cone'_N\subset\cone'_N$, and multiplication by the positive non-increasing sequence $1/d$ preserves $\cone'_N$.

Now let $\alpha(v):=\min_nv_n/y_n$ and maximize $\alpha$ over the compact set $S:=\{v\in\cone'_N:\ v\le y\}$. The set $S$ is mapped into itself by $Q$, since $Qv\le Qy=y$. The maximum $\alpha^*$ is positive, since a small constant vector lies in $S$. Let $v^*\in S$ attain it, so $v^*\ge\alpha^*y$. If $v^*\ne\alpha^*y$, then $Q(v^*-\alpha^*y)$ has all entries positive, so $\alpha(Qv^*)>\alpha^*$, a contradiction. Hence $y=v^*/\alpha^*$ is non-increasing, that is, $z\in\cone_N$.

\emph{$N=\infty$.} By Lemma~\ref{lem:basic}(d), $\kappa^{(N)}\to\kappa$. Let $\xi^N\in\cone_N$ be the unit eigenvectors above, padded by zeros, and choose a subsequence with $\xi^N\rightharpoonup\xi$ weakly. Since $K$ is compact, $\ip{\xi^N}{K\xi^N}\to\ip{\xi}{K\xi}$, and $\ip{\xi^N}{K\xi^N}=\kappa^{(N)}\to\kappa$; so $\ip{\xi}{K\xi}=\kappa>0$. Thus $\xi\ne0$, and as $\|\xi\|\le1$ its Rayleigh quotient is at least $\kappa$, which forces $\|\xi\|=1$ and $K\xi=\kappa\xi$. Weak convergence implies convergence of each coordinate, which preserves $\xi_n\ge0$ and the monotonicity of $y$. Finally $y_0>0$, since a non-zero non-negative non-increasing sequence has a positive first entry.
\end{proof}

\subsection{Proof of Theorem~\ref{thm:comp}}

\begin{proof}
Take $\xi$ from Lemma~\ref{lem:cone} for the kernel $a$. By~\eqref{eq:Mj} and Lemma~\ref{lem:mu},
\[
\kappa[b]\ \ge\ \ip{\xi}{K[b]\xi}=\kappa[a]+\sum_{j\ge1}(b_j-a_j)\,\mu_j(\xi)\ \ge\ \kappa[a]+(b_1-a_1)\,y_0^2\ >\ \kappa[a].
\]
The rearrangement is justified by $\sum_j(b_j-a_j)\|M_j\|<3\sum_j(b_j-a_j)<\infty$. For a truncation the same computation applies, with $\mu_j^{(N)}$ and the eigenvector of $K^{(N)}[a]$; there only $j\le2N-1$ occur.
\end{proof}

\begin{remark}\label{rem:derivative}
Where $h$ is differentiable, the same computation gives $\partial h/\partial(\vp^2)=\sum_jj^2(\vp^2+j^2)^{-2}\mu_j(\xi)>0$. KAY grouped this derivative by the weights $j^2/(\vp^2+j^2)^2$, which requires them to decrease in $j$; that holds if and only if $\vp\le\sqrt2$~\cite[Lemma~3]{KAY2025II}. Grouping by the transfer $j$ needs no condition on the weights.
\end{remark}

\section{Proofs of uniqueness, the sharp bound and Einstein optimality}\label{sec:proofcons}

We use one elementary estimate twice.

\begin{lemma}[Zero-temperature divergence]\label{lem:diverge}
Let $N\ge1$, $0\le\varepsilon\le1$, and let $a$ be a summable real sequence with $1-\varepsilon\le a_j\le1$ for $1\le j\le2N-1$. Then $\kappa[a]\ge\kappa^{(N)}[a]\ge2S_N-1-\varepsilon\,(2S_N+1)$, where $S_N:=\sum_{n<N}1/w_n$.
\end{lemma}

\begin{proof}
Test $K^{(N)}[a]$ with $\xi_n=1/\sqrt{w_n}$, $n<N$, that is $y_n=1/w_n$. The reference matrix $K^{(N)}[1]$ is defined directly by the finite formula~\eqref{eq:K}; its Rayleigh quotient at this vector is
\[
\frac{2\big(\sum_ny_n\big)^2-\sum_n y_n^2-\sum_n2n\,y_n^2}{\sum_n w_ny_n^2}=\frac{2S_N^2-S_N}{S_N}=2S_N-1 .
\]
Write $a_j=1-\varepsilon_j$ with $0\le\varepsilon_j\le\varepsilon$. The off-diagonal entries of $K^{(N)}[a]-K^{(N)}[1]$ are $-(\varepsilon_{|n-m|}+\varepsilon_{n+m+1})/\sqrt{w_nw_m}$, of absolute value at most $2\varepsilon/\sqrt{w_nw_m}$; the diagonal entries are $(2\sum_{j\le n}\varepsilon_j-\varepsilon_{2n+1})/w_n$, of absolute value at most $\varepsilon$. At the test vector the quadratic form of the difference is therefore at most $2\varepsilon S_N^2+\varepsilon S_N$ in absolute value, and dividing by $\|\xi\|^2=S_N$ gives the claim. The first inequality is Lemma~\ref{lem:basic}(d).
\end{proof}

Since $S_N\to\infty$, the right-hand side can be made as large as we like.

\begin{proof}[Proof of Theorem~\ref{thm:unique}]
\emph{Monotonicity.} If $\vp<\vp'$, then $\vp^2/(\vp^2+j^2)<\vp'^2/(\vp'^2+j^2)$ for every $j$, the differences are summable, and the Einstein kernels are kernels; Theorem~\ref{thm:comp} gives $h(\vp)<h(\vp')$. If $T'<T$, then $a_j(P,T')>a_j(P,T)$ for every $j$ because $P$ gives no mass to $\{0\}$, and $\sum_j a_j(P,T')<\infty$; so $k(P,T')>k(P,T)$.

\emph{Continuity.} By Lemma~\ref{lem:basic}(a) and Weyl's inequality, $|\kappa[a]-\kappa[b]|\le3\sum_j|a_j-b_j|$. For $T,T'\ge T_0>0$, each $|a_j(P,T)-a_j(P,T')|$ tends to $0$ as $T'\to T$ by dominated convergence, and is at most $2\langle\omega^2\rangle/(2\pi T_0j)^2$, which is summable in $j$. Hence $T\mapsto k(P,T)$ is continuous. The Einstein case is the special case $P=\delta_\Omega$.

\emph{Limits.} As $T\to\infty$, Theorem~\ref{thm:bound}(a), whose proof uses only Theorem~\ref{thm:comp}, gives $0<k(P,T)<g\langle\omega^2\rangle/(2\pi T)^2\to0$; likewise $h(\vp)<g\vp^2\to0$ as $\vp\to0$. As $T\to0$, $a_j(P,T)\to1$ for every $j$ by dominated convergence, so Lemma~\ref{lem:diverge} gives $\liminf_{T\to0}k(P,T)\ge2S_N-1$ for every $N$, and $S_N\to\infty$. The same argument gives $h(\vp)\to\infty$ as $\vp\to\infty$.

\emph{Conclusion.} Thus $T\mapsto k(P,T)$ is a decreasing homeomorphism of $(0,\infty)$ onto itself, and $T_c(\lambda,P)$ is the unique solution of $k(P,T)=1/\lambda$; it increases continuously and strictly with $\lambda$. For KAY-admissible $P$ the stability statement is~\cite[Thm.~1$^+$]{KAY2025II}. For Einstein phonons, $k(\delta_\Omega,T)=h(\Omega/2\pi T)$ gives Theorem~\ref{thm:unique}(d).

\emph{Truncations.} The same arguments, with the truncated version of Theorem~\ref{thm:comp} and the bound $\|K^{(N)}[a]-K^{(N)}[b]\|\le3\sum_j|a_j-b_j|$ (Lemma~\ref{lem:basic}(a) for the $N\times N$ blocks), prove the monotonicity and continuity of $h^{(N)}$ and $k^{(N)}$. Their limits are finite: $k^{(N)}(P,T)\to\kappa^{(N)}[(1)_j]$ as $T\to0$ (compare~\cite[Thm.~5]{KAY2025II}).
\end{proof}

\begin{proof}[Proof of Theorem~\ref{thm:bound}]
(a) For $\vp>0$ the Einstein kernel satisfies $\vp^2/(\vp^2+j^2)<\vp^2/j^2$ for every $j$, and Theorem~\ref{thm:comp} with $a=(\vp^2/(\vp^2+j^2))_j$ and $b=\vp^2(j^{-2})_j$ gives $h(\vp)<\kappa[\vp^2(j^{-2})]=\vp^2g$ by Lemma~\ref{lem:basic}(b). For general $P\in\Pt$,
\[
a_j(P,T)=\int\frac{\omega^2}{\omega^2+(2\pi Tj)^2}\,P(d\omega)<\frac{\langle\omega^2\rangle}{(2\pi T)^2}\,j^{-2}
\]
for every $j$, and the same argument gives $k(P,T)<g\langle\omega^2\rangle/(2\pi T)^2$.

(b) At $T=T_c$ we have $1=\lambda k(P,T_c)<\lambda g\langle\omega^2\rangle/(2\pi T_c)^2$, which rearranges to the claim; the Einstein case is $P=\delta_\Omega$.

(c) Let $b_j:=\langle\omega^2\rangle(2\pi T)^{-2}j^{-2}$. Then
\[
0<b_j-a_j(P,T)=\int\frac{\omega^4}{(2\pi Tj)^2\big(\omega^2+(2\pi Tj)^2\big)}\,P(d\omega)\le\frac{\langle\omega^4\rangle}{(2\pi T)^4\,j^4},
\]
so by Lemma~\ref{lem:basic}(a), $|k(P,T)-g\langle\omega^2\rangle/(2\pi T)^2|\le3\zeta(4)\langle\omega^4\rangle/(2\pi T)^4$. Dividing by $g\langle\omega^2\rangle/(2\pi T)^2$ shows that the ratio tends to $1$ as $T\to\infty$. As $\lambda\to\infty$, $T_c(\lambda,P)\to\infty$ by Theorem~\ref{thm:unique}, and $\lambda k(P,T_c)=1$ then gives $(2\pi T_c)^2/(g\lambda\langle\omega^2\rangle)\to1$. For Einstein phonons $\langle\omega^4\rangle=\Omega^4$.
\end{proof}

\begin{proof}[Proof of Theorem~\ref{thm:einstein}]
For fixed $T$ and $j$ the function $f_j(s)=s/(s+(2\pi Tj)^2)$ is strictly concave on $s\ge0$. By Jensen's inequality,
\[
a_j(P,T)=\int f_j(\omega^2)\,P(d\omega)\le f_j(\langle\omega^2\rangle)=\frac{\Omega_2^2}{\Omega_2^2+(2\pi Tj)^2},
\]
which is the Einstein kernel at frequency $\Omega_2$, i.e.\ at $\vp=\Omega_2/(2\pi T)$. Equality holds for some (equivalently every) $j$ only if $\omega^2$ is $P$-almost surely constant, i.e.\ $P=\delta_{\Omega_2}$. Otherwise the inequality is strict for every $j$, and Theorem~\ref{thm:comp} gives $k(P,T)<h(\Omega_2/2\pi T)$. At $T=T_c(\lambda,P)$ this gives $\lambda h(\Omega_2/2\pi T)>1$, so by Theorem~\ref{thm:unique} the Einstein system at frequency $\Omega_2$ is still below its critical temperature: $T_c(\lambda,P)<T_c(\lambda,\Omega_2)$.
\end{proof}

\begin{lemma}[Convexity in the spectrum]\label{lem:jensen}
For every $P\in\Pt$ and $T>0$, $k(P,T)\le\int h\big(\omega/(2\pi T)\big)\,P(d\omega)$.
\end{lemma}

\begin{proof}
Write $e(\vp):=\big(\vp^2/(\vp^2+j^2)\big)_{j\ge1}$ for the Einstein kernel, so that $a_j(P,T)=\int e_j\big(\omega/(2\pi T)\big)\,P(d\omega)$ and $h(\vp)=\kappa[e(\vp)]$. The function $h$ is continuous (Theorem~\ref{thm:unique}; the continuity part of its proof uses only Lemma~\ref{lem:basic}(a)) and satisfies $0<h(\vp)\le3\sum_je_j(\vp)\le3\pi\vp/2$ by Lemma~\ref{lem:basic}(a) with $b=0$ and $\sum_{j\ge1}\vp^2/(\vp^2+j^2)\le\int_0^\infty\vp^2/(\vp^2+x^2)\,dx$; since $P\in\Pt$ has a finite first moment, the integral is finite. Let $\xi$ have finitely many non-zero entries. Then $\ip{\xi}{K[a]\xi}$ is a finite linear combination of the $a_j$, so
\[
\ip{\xi}{K[a(P,T)]\xi}=\int\ip{\xi}{K[e(\omega/2\pi T)]\xi}\,P(d\omega)\le\|\xi\|^2\int h\big(\omega/(2\pi T)\big)\,P(d\omega).
\]
Such vectors are dense in $\ell^2$ and $K[a(P,T)]$ is bounded, so taking the supremum of the Rayleigh quotient over them gives the claim.
\end{proof}

\begin{proof}[Proof of Theorem~\ref{thm:moments}]
For $\vp\le1$, $h(\vp)/\vp^{n+1}<g\vp^{1-n}\le g$ by Theorem~\ref{thm:bound}(a), and for $\vp\ge1$, $h(\vp)/\vp^{n+1}\le(3\pi/2)\vp^{-n}\le3\pi/2$ by the estimate in the proof of Lemma~\ref{lem:jensen}. Theorem~\ref{thm:bound}(a) and~(c) give $H_1=g$. By Lemma~\ref{lem:jensen} and $h(\vp)\le H_n\vp^{n+1}$,
\[
\lambda k(P,T)\le\lambda H_n\int\Big(\frac{\omega}{2\pi T}\Big)^{n+1}P(d\omega)=\frac{2H_nM_n}{(2\pi T)^{n+1}} .
\]
At $T=T_c(\lambda,P)$ the left-hand side equals $1$ by Theorem~\ref{thm:unique}, which gives the bound on $T_c$. For sharpness, fix $M_n>0$ and $\vp>0$, put $T_\vp:=\frac{1}{2\pi}\big(2M_nh(\vp)/\vp^{n+1}\big)^{1/(n+1)}$ and $\Omega:=2\pi T_\vp\vp$, and take the Einstein spectrum $\alpha^2F=M_n\Omega^{-n}\delta_\Omega$, whose $n$-th moment is $M_n$ and whose coupling is $\lambda=2M_n\Omega^{-n-1}$. Then $\lambda h(\Omega/2\pi T_\vp)=2M_nh(\vp)/(2\pi T_\vp\vp)^{n+1}=1$, so $T_\vp=T_c(\lambda,\Omega)$ by Theorem~\ref{thm:unique}(d). The supremum of $T_\vp$ over $\vp>0$ is $\frac{1}{2\pi}(2H_nM_n)^{1/(n+1)}$.
\end{proof}

\section{Proof of Theorem~\ref{thm:conj3}}\label{sec:proofconj3}

The proof decomposes a non-increasing vector into the prefix indicators $u_l$ of Section~\ref{sec:proofcomp} and has three steps: the form on a single prefix, the cross terms between nested prefixes, and the strict decrease of $c_l[a]$ in $l$.

\begin{proof}
Write $c_l:=c_l[a]$. For $1\le l\le N$ let $u_l\in\RR^N$ be the indicator of $\{0,\dots,l-1\}$, and let $\mathcal B$ be the symmetric bilinear form of $\mathcal G_N[a]-c_N\mathcal D_N$.

\emph{Prefixes.} On a single prefix the diagonal subtraction cancels the Toeplitz part exactly, because $\sum_{n\ne m<l}a_{|n-m|}=2\sum_{n<l}A_n$. Hence $\mathcal G_N[a](u_l)=\sum_{n,m<l}a_{n+m+1}=l^2c_l$, and $\mathcal D_N(u_l)=\sum_{n<l}w_n=l^2$. So $\mathcal B(u_l,u_l)=l^2(c_l-c_N)$.

\emph{Cross terms.} For $l\le l'$, the vector $u_{l'}-u_l$ is supported on $\{l,\dots,l'-1\}$, disjoint from the support of $u_l$. All matrix entries of $\mathcal G_N$ that couple the two supports are off-diagonal and positive, and $\mathcal D_N$ is diagonal. Therefore $\mathcal B(u_l,u_{l'})\ge\mathcal B(u_l,u_l)=l^2(c_l-c_N)$.

\emph{$c_l$ is strictly decreasing.} Let $U,V$ be independent and uniform on $[0,1)$. Then $\lfloor lU\rfloor$ is uniform on $\{0,\dots,l-1\}$, so $c_l=\mathbb E\,a_{\lfloor lU\rfloor+\lfloor lV\rfloor+1}$. For $l<l'$, $\lfloor lU\rfloor\le\lfloor l'U\rfloor$ pointwise, with strict inequality when $U>1-1/l'$. Since $a$ is strictly decreasing, $c_l>c_{l'}$.

\emph{Conclusion.} Every non-negative non-increasing $\theta$ is $\theta=\sum_{l=1}^Nx_lu_l$ with $x_l=\theta_{l-1}-\theta_l\ge0$ ($\theta_N:=0$). Then
\[
\mathcal B(\theta,\theta)=\sum_{l,l'}x_lx_{l'}\,\mathcal B(u_l,u_{l'})\ \ge\ \sum_{l,l'}x_lx_{l'}\,\min\{l,l'\}^2\,\big(c_{\min\{l,l'\}}-c_N\big)\ \ge\ 0 .
\]
The diagonal terms with $l<N$ are $x_l^2\,l^2(c_l-c_N)$, which are positive unless $x_l=0$. So equality forces $x_l=0$ for all $l<N$, i.e.\ $\theta$ constant. Conversely, a constant $\theta=x_Nu_N$ gives $\mathcal B(\theta,\theta)=x_N^2\,\mathcal B(u_N,u_N)=0$.
\end{proof}

\section{Machine-checked and computer-assisted verification}\label{sec:lean}

\subsection{Lean formalization}

The finite-dimensional core of the argument is formalized in Lean~4 (v4.35.0) with Mathlib~\cite{Lean4,Mathlib2020}. For $N\ge1$ and an arbitrary real sequence $a$, the development defines the matrix $K^{(N)}[a]$ of~\eqref{eq:K} on $\mathrm{Fin}\,N$ entry by entry, its quadratic form, and $\kappa^{(N)}[a]$ as the supremum of the quadratic form over the unit sphere. A kernel (\lean{IsKernel}) is a sequence that is positive and non-increasing for $j\ge1$; summability plays no role at finite $N$. The cone \lean{InCone} is $\cone_N$. Discrete phonon spectra are finite families of frequencies $\omega_i>0$ with non-negative weights summing to one. Table~\ref{tab:lean} lists the formalized theorems. Each depends only on Lean's three standard axioms (\lean{propext}, \lean{Classical.choice}, \lean{Quot.sound}), as reported by \lean{\#print axioms}; the development contains no \lean{sorry}, no added axioms and no \lean{native\_decide}.

\begin{table}[t]
\caption{Lean theorems and the statements they formalize, all for the truncation to $N\ge1$ frequencies.}
\label{tab:lean}
\begin{tabular}{@{}ll@{}}
\toprule
Lean name & Statement\\
\midrule
\lean{exists\_max}, \lean{qf\_le\_kappa} & $\kappa^{(N)}[a]$ is attained on the unit sphere\\
\lean{exists\_cone\_maximizer} & Lemma~\ref{lem:cone} for finite $N$: a maximizer lies in $\cone_N$\\
\lean{qf\_sub}, \lean{mu\_nonneg}, \lean{mu\_one\_ge} & decomposition~\eqref{eq:Mj} and Lemma~\ref{lem:mu}\\
\lean{comparison} & Theorem~\ref{thm:comp} for $\kappa^{(N)}$\\
\lean{kappa\_pos} & $\kappa^{(N)}[a]\ge a_1$ for a kernel $a$, Lemma~\ref{lem:basic}(c)\\
\lean{kappa\_smul} & $\kappa^{(N)}[sa]=s\,\kappa^{(N)}[a]$ for $s>0$\\
\lean{einstein\_lt} & $h^{(N)}(\vp)<\vp^2\,\kappa^{(N)}[(j^{-2})]$ for $\vp>0$, Theorem~\ref{thm:bound}(a)\\
\lean{einstein\_strictMonoOn} & $\vp\mapsto h^{(N)}(\vp)$ is strictly increasing on $\vp>0$\\
\lean{disp\_lt} & Theorem~\ref{thm:bound}(a) for discrete spectra\\
\lean{disp\_strictAntiOn} & $T\mapsto k^{(N)}(P,T)$ is strictly decreasing on $T>0$, discrete $P$\\
\lean{disp\_le\_einstein} & $k^{(N)}(P,T)\le h^{(N)}(\Omega_2/2\pi T)$ for $T>0$, discrete $P$\\
\lean{conj3}, \lean{conj3\_eq\_iff} & Theorem~\ref{thm:conj3}, inequality and equality case\\
\botrule
\end{tabular}
\end{table}

\emph{What is not formalized.} (i) The passage $N\to\infty$: Lemma~\ref{lem:basic}(d) and the case $N=\infty$ of Lemma~\ref{lem:cone} (its first assertion); (ii) continuity and the limits in Theorem~\ref{thm:unique}, and the construction and properties of $T_c$ in its parts (c)--(d); (iii) the conversion of the spectral bound to a $T_c$ bound in Theorem~\ref{thm:bound}(b), and the sharpness assertion in its part (c); (iv) spectra that are not finite sums of point masses, and the strict form of Theorem~\ref{thm:einstein}; (v) the derivation of Section~\ref{sec:setting}, Lemma~\ref{lem:jensen}, Theorem~\ref{thm:moments} and Corollary~\ref{cor:coulomb}; (vi) KAY's identification of the instability with $\lambda k>1$. The steps in (i)--(v) use standard facts about compact operators, dominated convergence and Jensen's inequality and are proved in full in the text; step (vi) is the cited result of KAY~\cite[Thm.~1$^+$]{KAY2025II} and is not reproved here. In particular, the formalized Einstein bound has the truncated constant $\kappa^{(N)}[(j^{-2})]\le g$; the statement $h(\vp)<g\vp^2$ for the full operator needs the case $N=\infty$ of Lemma~\ref{lem:cone}.

\subsection{Computer-assisted proofs}

Before the comparison principle was found, the Einstein inequality $h(\vp)<g\vp^2$ was proved using ball arithmetic. A second implementation, written from the list of lemmas alone, proves it in exact rational arithmetic. Neither uses Theorem~\ref{thm:comp}; together they provide independent checks of Theorem~\ref{thm:bound} for Einstein phonons. Lemma~\ref{lem:jensen} does not use Theorem~\ref{thm:comp} either, and with $h(\vp)<g\vp^2$ it gives $\lambda k(P,T)<g\lambda\langle\omega^2\rangle/(2\pi T)^2$ for every $P\in\Pt$. So the certificates also prove, independently of the comparison principle, that $\lambda k(P,T)<1$ for every $T\ge\frac{1}{2\pi}\sqrt{g\lambda\langle\omega^2\rangle}$: the normal state is linearly stable above the bound for every spectrum. Appendix~\ref{app:cap} gives the details.

\subsection{Use of large language models}\label{sec:ai}

This research used large language models (LLMs) extensively, and we describe their role as the journal's policy requires. The author selected the problem, directed the research throughout, decided which approaches to pursue and which to abandon, reviewed the statements and proofs of all results in full, and is responsible for the content of this paper. LLM-based coding agents (Anthropic Claude and OpenAI GPT models, used through an agent harness) assisted, under the author's direction, with the literature search, the exploration of proof strategies and the drafting of proofs, the certificate programs and the Lean formalization, adversarial audits of the arguments by models of different families, and drafting and editing the text. No output was accepted on the authority of a model. Every new result in the paper is proved in the text and checked by the author; its finite-dimensional core is in addition machine-checked in Lean, and the Einstein bound is confirmed by the certificates of Appendix~\ref{app:cap}. No LLM is an author of this paper.

\section{Scope and open problems}\label{sec:scope}

Apart from Corollary~\ref{cor:coulomb}, all results are theorems about the linearized, isotropic Eliashberg model with $\mu^*=0$, in the formulation of KAY~\cite{KAY2025II,KAY2025III}, where the critical temperature is the linear-stability boundary of the normal state. They do not address the nonlinear gap equations or the order of the transition, a Coulomb pseudopotential $\mu^*>0$ beyond the bound of Corollary~\ref{cor:coulomb}, anisotropic or multiband pairing, vertex corrections, or the breakdown of Migdal--Eliashberg theory at large $\lambda$~\cite{YuzbashyanAltshuler2022Breakdown,EsterlisBreakdown2018}. That breakdown motivates bounds on $T_c$ of a different kind~\cite{Esterlis2018}. The bound is a statement about this model, not a prediction for any material.

Four questions seem natural. (i) With $\mu^*>0$ the kernel acquires a negative, frequency-independent part and the cone argument does not apply as it stands. Corollary~\ref{cor:coulomb} bounds every instability temperature, but whether that temperature is unique, and whether a comparison principle survives, is open. (ii) The comparison principle concerns the linear problem. Whether the full nonlinear gap equations have a non-trivial solution exactly for $T<T_c$, so that the linear and thermodynamic transition temperatures coincide for all $\lambda$, is part of KAY's broader programme~\cite{KAY2025II}. (iii) The cone $\cone$ seems to be the natural home of physical gap functions in this model. Other monotonicity questions, for instance in the $\gamma$-model with general $\gamma$~\cite{KAY2025I,ElezabyAbanov2026}, may yield to the same decomposition by transfer. (iv) Theorem~\ref{thm:moments} leaves the value of $H_0$ uncertified, which would need an upper bound on $h(\vp)/\vp$ uniformly for large $\vp$, and the moments $-1<n<0$ open.

\backmatter

\bmhead{Supplementary information}
The supplementary archive contains the Lean project (source and build instructions), the two certificate programs with their receipts, and the scripts that produce the figures, the values quoted in Appendix~\ref{app:cap} and the floating-point value of $H_0$ after Theorem~\ref{thm:moments}.

\section*{Declarations}

\bmhead{Funding}
This work was carried out at, and funded by, SuperMatics Inc. It received no external funding.

\bmhead{Competing interests}
The author is a founder of SuperMatics Inc., which develops computational methods for materials discovery. The author declares no other competing interests.

\bmhead{Ethics approval and consent to participate}
Not applicable.

\bmhead{Consent for publication}
Not applicable.

\bmhead{Data availability}
No experimental data were generated. The supplementary archive contains the programs that produce the paper's own numerical results, together with the certificate receipts.

\bmhead{Materials availability}
Not applicable.

\bmhead{Code availability}
The Lean formalization, the certificate programs and the figure scripts are included in the supplementary archive, which is also distributed as the ancillary files of the arXiv version of this paper.

\bmhead{Author contribution}
D.R.\ selected the problem, directed the research, reviewed and checked all statements and proofs, and wrote the paper, with the assistance of large language models described in Section~\ref{sec:ai}.

\begin{appendices}
\renewcommand{\theHtable}{\thesection.\arabic{table}}

\section{Computer-assisted proofs of the Einstein inequality}\label{app:cap}

We prove $\Sp G(\vp)<g$ for every $\vp>0$, where $c_j(\vp):=1/(\vp^2+j^2)$ and $G(\vp):=K[c(\vp)]$; by Lemma~\ref{lem:basic}(b), $h(\vp)=\vp^2\,\Sp G(\vp)$, so this is Theorem~\ref{thm:bound}(a) for Einstein phonons. Note that $G_0=G(0)$. Decompose $G(\vp)=P(\vp)-D(\vp)$ with $P(\vp)=\sum_jc_j(\vp)(T_j+H_j)\ge0$ entrywise and $D(\vp)=\sum_jc_j(\vp)D_j\ge0$ diagonal, and let $S(\vp)=\sum_{k\ge1}c_k(\vp)=\frac{\pi \vp\coth(\pi \vp)-1}{2\vp^2}$, a decreasing function.

\begin{lemma}\label{lem:tools}\leavevmode
\begin{enumerate}
\item[(a)] (Envelope.) If $0\le \vp_l\le \vp\le \vp_u$, then $\Sp G(\vp)\le\Sp U$ with $U:=P(\vp_l)-D(\vp_u)$.
\item[(b)] (Block.) If $U=\begin{psmallmatrix}A&B\\B^{\mathsf T}&C\end{psmallmatrix}$ with $\Sp A\le\alpha$, $\Sp C\le\gamma$ and $\|B\|\le\beta$, then $\Sp U\le\frac{\alpha+\gamma}{2}+\sqrt{(\frac{\alpha-\gamma}{2})^2+\beta^2}$.
\item[(c)] (Cholesky certificate.) If $L$ is an exactly representable lower-triangular matrix and $\tau>0$ with $\|(sI-A)-\tau I-LL^{\mathsf T}\|_F<\tau$, then $\Sp A<s$.
\item[(d)] (Tails.) On the coordinates $n\ge N$, $\Sp C\le3S(\vp_l)/w_N$, and
\begin{gather*}
\|B\|_F^2\le\sum_{n<N}\frac{2}{w_nw_N}\big[T_4(N-n)+T_4(N+n+1)\big],\\
T_4(m):=m^{-4}+\tfrac{1}{3m^3}\ge\sum_{k\ge m}k^{-4}.
\end{gather*}
\item[(e)] (Large $\vp$.) $\Sp G(\vp)\le\|P(\vp)\|\le3S(\vp)$.
\end{enumerate}
\end{lemma}

\begin{proof}
(a) $\ip{x}{G(\vp)x}\le\ip{|x|}{P(\vp)|x|}-\ip{|x|}{D(\vp)|x|}\le\ip{|x|}{U|x|}$, because each $c_j$ decreases in $\vp$. (b) $\ip{x}{Ux}\le\alpha\|x_1\|^2+2\beta\|x_1\|\|x_2\|+\gamma\|x_2\|^2$. (c) $sI-A=LL^{\mathsf T}+\tau I+E$ with $\|E\|\le\|E\|_F<\tau$. (d) $C$ is bounded above by the tail of $P(\vp_l)$; the Schur test with weights $w_n^{-1/2}$ bounds its norm by $\sup_{n\ge N}\sum_{m\ge N}[(1-\delta_{nm})c_{|n-m|}+c_{n+m+1}]/w_m\le3S/w_N$. The Frobenius bound uses $c_k\le k^{-2}$ and $w_m\ge w_N$. (e) The same Schur test over all indices.
\end{proof}

\paragraph{Primary certificate (ball arithmetic).} A floating-point eigenvector of the $N=256$ truncation of $G_0$, rounded to an exact dyadic vector and then normalized in Arb to a unit vector $\tilde\psi$, gives the rigorous lower bound $q_0\ge\QzeroLow$ on $g$ and the residual $\rho_0\le\RhoZero$. By min--max and Lemma~\ref{lem:tools}(b)--(d) applied to $G_0-t\tilde\psi\tilde\psi^{\mathsf T}$, the second eigenvalue of $G_0$ is at most $g_2\le\GtwoUp$ (its floating-point value is $\LamTwoFloat$), and the Davis--Kahan theorem~\cite{DavisKahan1970} bounds the distance of $\tilde\psi$ to the unit top eigenvector $\psi$ of $G_0$, with its sign chosen so that $\ip{\psi}{\tilde\psi}\ge0$, by $\delta\le\sqrt2\rho_0/(q_0-g_2)$. The half-line $\vp>0$ is split into three regions.
\begin{itemize}
\item \emph{Region I}, $0<\vp\le1/4$. Here $G_0-G(\vp)=\vp^2R(\vp)$ with $R(\vp)=\sum_j\frac{M_j}{j^2(\vp^2+j^2)}$ and $\|R(\vp)\|\le\rho\le\RegOneRho$. For a unit vector $x=a\psi+b\phi$ with $\phi\perp\psi$ and $\|\phi\|=1$, $\ip{x}{(g-G(\vp))x}\ge(g-g_2)b^2+\vp^2(ra^2-2\rho|a||b|-\rho b^2)$ with $r=\inf_\vp\ip{\psi}{R(\vp)\psi}\ge\RegOneR$, and the resulting positivity criterion $r(q_0-g_2-\vp_a^2\rho)-\vp_a^2\rho^2$, with $\vp_a=1/4$, is at least $\RegOneCrit>0$.
\item \emph{Region II}, $1/4\le \vp\le4$, is divided into $\RegTwoCells$ cells with dyadic endpoints, obtained by rounding a geometric grid of ratio $\RegTwoRatio$. In each cell, Lemma~\ref{lem:tools}(a)--(d) with $N=\RegTwoN$ and a Cholesky certificate bound $\Sp G(\vp)$ below $q_0$. The smallest relative margin is $\RegTwoWorstMargin\,\%$, on $[\RegTwoWorstLo,\RegTwoWorstHi]$.
\item \emph{Region III}, $\vp\ge\RegThreeVb$, is controlled by $\Sp G(\vp)\le3S(\RegThreeVb)\le\RegThreeVal<q_0$.
\end{itemize}
All bounds are evaluated in Arb ball arithmetic~\cite{Johansson2017Arb} at $\ArbPrec$ bits, through python-flint~$\FlintVer$~\cite{PythonFlint} on FLINT~3.6.0~\cite{FLINT}. Floating point only proposes trial objects (the unnormalized trial vector and the Cholesky factors $L$), which are then treated as exact dyadic data; the trial vector is normalized in Arb. The run takes about $\CertRuntime$\,s on a laptop. Table~\ref{tab:cap} lists the certified quantities.

\paragraph{Independent certificate (exact rationals).} A second implementation works after the congruence $W^{1/2}(\cdot)W^{1/2}$, $W=\operatorname{diag}(w_n)$, in which all entries are rational, and checks every inequality in exact rational arithmetic (FLINT \texttt{fmpq}). It uses $N=\IndepN$, the regions $(0,\IndepVa]$, $[\IndepVa,\IndepVb]$ ($\IndepCells$ geometric cells of ratio $\IndepRatio$) and $[\IndepVb,\infty)$, the rational lower bound $g\ge\IndepQ$, and rational bounds on the integrals in place of $\coth$. It certifies $g-\Sp G(\vp)\ge\IndepKappa\,\vp^2$ on the first region, a gap of at least $\IndepMidGap$ below $\IndepQ$ on the second and of at least $\IndepLargeGap$ on the third, in $\IndepRuntime$\,s.

\begin{table}[t]
\caption{Bounds from the receipt of the primary certificate.}
\label{tab:cap}
\begin{tabular}{@{}lll@{}}
\toprule
Quantity & Certified bound & Role\\
\midrule
$q_0$ (Rayleigh quotient, $N=256$) & $\ge\QzeroLow$ & lower bound on $g$\\
$\rho_0$ (residual) & $\le\RhoZero$ & eigenvector error\\
$g_2$ (second eigenvalue of $G_0$) & $\le\GtwoUp$ & spectral gap\\
$\rho$ ($\sup_\vp\|R(\vp)\|$) & $\le\RegOneRho$ & Region I\\
$r$ ($\inf_\vp\ip{\psi}{R(\vp)\psi}$) & $\ge\RegOneR$ & Region I\\
Region I criterion & $\RegOneCrit>0$ & Region I\\
Region II worst margin & $\RegTwoWorstMargin\,\%$ ($\RegTwoCells$ cells) & Region II\\
$3S(\RegThreeVb)$ & $\le\RegThreeVal$ & Region III\\
\botrule
\end{tabular}
\end{table}

\paragraph{Provenance.} The plan of the primary certificate was fixed before it was run (SHA-256 \hash{\PlanSha}). Primary certificate script \hash{\CertSha}, receipt \hash{\ReceiptSha}; independent certificate script \hash{\IndepScriptSha}, receipt \hash{\IndepSha}.

\end{appendices}

\bibliography{refs}

\end{document}